\documentclass[conference]{IEEEtran}

\usepackage{cite}
\usepackage{amsmath}
\usepackage{array}
\usepackage{graphicx}
\usepackage{url}

\usepackage{color}

\newcommand{\HT}{\mathsf{H}}

\begin{document}
\bstctlcite{IEEEexample:BSTcontrol}

\title{Maximum likelihood convolutional beamformer\\ for simultaneous denoising and dereverberation}

\author{Tomohiro~Nakatani and Keisuke~Kinoshita\\
NTT Communication Science Laboratories, NTT Corporation, Japan}

\markboth{}{}

\maketitle

\begin{abstract}
This article describes a probabilistic formulation of a Weighted Power minimization Distortionless response convolutional beamformer (WPD).  The WPD unifies a weighted prediction error based dereverberation method (WPE) and a minimum power distortionless response beamformer (MPDR)  into a single convolutional beamformer, and achieves simultaneous dereverberation and denoising in an optimal way.  However, the optimization criterion is obtained simply by combining existing criteria without any clear theoretical justification.  This article presents a generative model and a probabilistic formulation of a WPD, and derives an optimization algorithm based on a maximum likelihood estimation.  We also describe a method for estimating the steering vector of the desired signal by utilizing WPE within the WPD framework to provide an effective and efficient beamformer for denoising and dereverberation.
\end{abstract}


\IEEEpeerreviewmaketitle

\section{Introduction}
When a speech signal is captured by distant microphones, e.g., in a conference room, it will inevitably contain additive noise and reverberation components. These components are detrimental to the perceived quality of the observed speech signal and often cause serious degradation in many applications such as hands-free teleconferencing and ASR.

Microphone array signal processing has been investigated to minimize the aforementioned detrimental effects on the acquired signal.  A filter-and-sum beamformer \cite{Anguera07ASLP}, a minimum-variance distortionless response beamformer (MVDR) and MPDR \cite{MPDR,Cox,cgmmmvdr,erdogan16intersp,L1MPDR}, and a maximum signal-to-noise ratio beamformer \cite{GEV,maxSNR,BeamNET} are widely-used techniques for denoising, while WPE and its variants \cite{wpe,gwpe,owpe,swpe,cwpe} are emerging dereverberation techniques.  Techniques for reducing both noise and reverberation have also been investigated, for example, using both MVDR and WPE by cascade integration \cite{delcroix15eurasip,mvdrwpe,Togami,lukas2018interspeech}.  The usefulness of these techniques, particularly for ASR, has been extensively studied, e.g., at the REVERB challenge \cite{REVERB} and the CHiME-3/4/5 challenges \cite{CHiME3,CHiME4,CHiME5}. Moreover, advances in these techniques have led to recent progress on commercial devices, such as smart speakers \cite{GoogleHome,homepod,SPM}.

Recently, a new unified beamforming approach has been proposed for achieving denoising and dereverberation both simultaneously and optimally \cite{wpdarxiv,wpegsc}.  In \cite{wpdarxiv}, researchers introduced a convolutional beamformer that unifies WPE and MPDR, and presented a method for optimizing the beamformer by using a single optimization criterion. It is referred to as a Weighted Power minimization Distortionless response beamformer (WPD).  We showed experimentally that this beamformer provides substantially better simultaneous denoising and dereverberation performance than conventional cascade integration approaches\footnote{A recent study has revealed that the WPD yields the same outputs even when it is implemented with a certain cascade configuration consisting of a WPE and an MPDR. The details will be discussed in future publications.  The analysis presented in this paper is based on a form of convolutional beamformer that unifies a WPE and an MPDR into a single filter.}.
However, the reason for the effectiveness of the unified approach remains somewhat unclear because the optimization criterion was constructed simply by combining existing criteria without any clear theoretical justification.

To clarify the mechanism of the unified approach, this article presents a generative model and a probabilistic formulation of a WPD, and derives an optimization algorithm based on a maximum likelihood (ML) estimation.  We also present an iterative estimation method based on WPD for the steering vector of the desired signal, which was conventionally assumed to be given or estimated in advance. A key to successful estimation is to use multi-input multi-output (MIMO) dereverberation by WPE \cite{gwpe} within the WPD framework.  Experiments using the REVERB challenge dataset show the importance of using the MIMO dereverberation to achieve superior denoising and dereverberation in terms of objective speech enhancement measures and ASR performance.

In the remainder of this paper, we define the model of the signal and the beamformer in Section II, and present the probabilistic formulation of WPD in Section III.  Section IV derives the optimization algorithm and describes a processing flow that utilizes WPE.  The experimental results and concluding remarks are given in Sections V and VI, respectively.

\section{Model of signal and beamformer}
We assume that a single speech signal is captured by $M$ microphones in a noisy reverberant environment. The captured signal in the short-time Fourier transform (STFT) domain is approximately modeled at each frequency bin \cite{Nak-ICASSP2008} by 
\begin{equation}
\mathbf{x}_t=\sum_{\tau^=0}^{L_a} \mathbf{a}_{\tau}s_{t-\tau}+\mathbf{n}_t,
\label{eq:obs}
\end{equation}
where $t$ and $\tau$ are frame indices, $\mathbf{x}_t=[x_t^{(1)},x_t^{(2)},\ldots,x_t^{(M)}]^\top$ is, letting $\top$ denotes a non-conjugate transpose, a column vector containing STFT coefficients of all the microphone signals at a time frame $t$, $s_t$ is the STFT coefficient of the clean speech signal, $\mathbf{a}_t=[a_t^{(1)},a_t^{(2)},\ldots,a_t^{(M)}]^\top$ for $t=0,1,\ldots,L_a$ is a set of column vectors containing convolutional acoustic transfer functions from the speaker location to all the microphones, and $\mathbf{n}_t=[n_t^{(1)},n_t^{(2)},\ldots,n_t^{(M)}]^\top$ is the additive noise.  In this paper, the frequency indices of the symbols are omitted for brevity, and on the assumption that each frequency bin is processed independently in the same way. Hereafter, we refer to the STFT coefficients of signals simply as signals.

The first term in Eq.~(\ref{eq:obs}) can be further decomposed into two parts, one consisting of the direct signal and early reflections, and referred to as a desired signal $\mathbf{d}_t$, and the other corresponding to the late reverberation $\mathbf{r}_t$ \cite{early}. With this decomposition, Eq.~(\ref{eq:obs}) is rewritten as
\begin{align}
\mathbf{x}_t&=\mathbf{d}_t+\mathbf{r}_t+\mathbf{n}_t,\label{eq:obs2}\\
\mathbf{d}_t&=\sum_{\tau^=0}^{b-1} \mathbf{a}_{\tau}s_{t-\tau},\label{eq:desired}\\
\mathbf{r}_t&=\sum_{\tau^=b}^{L_a} \mathbf{a}_{\tau}s_{t-\tau},
\end{align}
where $b$ is a frame index that divides the reverberation into the two parts.  The goal of the beamforming is to preserve $\mathbf{d}_t$, while reducing $\mathbf{r}_t$ and $\mathbf{n}_t$ from $\mathbf{x}_t$.
In this paper, $\mathbf{d}_t$, $\mathbf{r}_t$, and $\mathbf{n}_t$ are assumed to be statistically independent of each other.  (See \cite{wpe} for a more precise discussion of the statistical independence of $\mathbf{d}_t$ and $\mathbf{r}_t$.)

With WPD, we further assume that the transfer function corresponding to the desired signal can be approximated by a product of a vector $\mathbf{v}$ with a clean speech signal, i.e., $\mathbf{d}_t=\mathbf{v}s_t$, in the STFT domain. Then Eq.~(\ref{eq:obs2}) becomes
\begin{equation}
  \mathbf{x}_t=\mathbf{v}s_t+\mathbf{r}_t+\mathbf{n}_t,\label{eq:obs2.5}\\
\end{equation}
Here, $\mathbf{v}$ is also termed a steering vector.  This paper sets $m=1$ as the reference microphone, and describes a method for estimating the desired signal at the microphone without loss of generality.  The desired signal at the reference microphone is represented as
\begin{equation}
  d_t^{(1)}=v^{(1)}s_t,
\end{equation}
where $v^{(1)}$ is the element of $\mathbf{v}$ at the reference microphone.

\subsection{Model of convolutional beamformer}
We now define a MIMO convolutional beamformer as
\begin{align}
  \mathbf{y}_t&={W}_0^{\HT}\mathbf{x}_t+\sum_{\tau=b}^{L_w}W_{\tau}^{\HT}\mathbf{x}_{t-\tau},\label{eq:bf}
\end{align}
where $\mathbf{y}_t$ is the output of the beamformer, $W_t$ for each $t$ $(=0,b,b+1,\ldots,L_w)$  is an $M\times M$ dimensional matrix, which is composed of the beamformer coefficients, $\HT$ denotes conjugate transpose, and $b$ is the prediction delay that corresponds to $b$ in Eq.~(\ref{eq:desired}) and is introduced to prevent the desired signal from being distorted by the convolutional beamforming \cite{wpe}.  We further decompose $W_t$ for each $t$ as follows
\begin{align}
  W_t&=[\mathbf{w}_t,B_t],\label{eq:Wt}
\end{align}
where $\mathbf{w}_t=[w_t^{(1)},w_t^{(2)},\ldots,w_t^{(M)}]^\top$ is an $M$-dimensional column vector that denoises and dereverberates the captured signal, and $B_t$ is an $M\times(M-1)$ dimensional matrix that extracts the noise from the captured signal. Note that our interest is in the estimation of $\mathbf{w}_t$, but $B_t$ is introduced because it is necessary for the probabilistic formulation of a WPD.  

To characterize $\mathbf{w}_t$ and $B_t$, we introduce the following constraints.
\begin{align}
  \mathbf{w}_0^{\HT}\mathbf{v}&=v^{(1)},\label{eq:dr}\\
  B_0^{\HT}\mathbf{v}&=0.\label{eq:orth}
\end{align}
Eq.~(\ref{eq:dr}) specifies that $\mathbf{w}_0$ extracts the desired signal at the reference channel with no distortion. Eq.~(\ref{eq:orth}) specifies that $B_0$ blocks the signal subspace spanned by $\mathbf{v}$ that includes the desired signal.  With the constraints, and based on Eqs.~(\ref{eq:obs2}) and (\ref{eq:obs2.5}), Eq.~(\ref{eq:bf}) can be rewritten as
\begin{align}
  \mathbf{y}_t&=\left[\begin{array}{c}y_t^{(1)}\\\mathbf{y}_t^{(2:M)}\end{array}\right]
  =\left[\begin{array}{c}d_t^{(1)}+\tilde{r}_t+\tilde{n}_t\\
      \dot{\mathbf{r}}_t+\dot{\mathbf{n}}_t\end{array}\right],\label{eq:output2}
\end{align}
where $y_t^{(1)}$ is the first element of $\mathbf{y}_t$, $\mathbf{y}_t^{(2:M)}$ is a column vector containing the other elements of $\mathbf{y}_t$, and
\begin{align}
  \tilde{r}_t&=\mathbf{w}_0^{\HT}\mathbf{r}_t+\sum_{\tau=b}^{L_w}\mathbf{w}_\tau^{\HT}(\mathbf{d}_{t-\tau}+\mathbf{r}_{t-\tau}),\\
  \tilde{n}_t&=\mathbf{w}_0^{\HT}\mathbf{n}_t+\sum_{\tau=b}^{L_w}\mathbf{w}_\tau^{\HT}\mathbf{n}_{t-\tau},\\
  \dot{\mathbf{r}}_t&=B_0^{\HT}\mathbf{r}_t+\sum_{\tau=b}^{L_w}B_\tau^{\HT}(\mathbf{d}_{t-\tau}+\mathbf{r}_{t-\tau}),\\
  \dot{\mathbf{n}}_t&=B_0^{\HT}\mathbf{n}_t+\sum_{\tau=b}^{L_w}B_\tau^{\HT}\mathbf{n}_{t-\tau},
\end{align}
In Eq.~(\ref{eq:output2}), $\tilde{r}_t$ and $\dot{\mathbf{r}}_t$ are the reverberation that remains after the beamforming, and $\tilde{n}_t$ and $\dot{\mathbf{n}}_t$ represent the noise that remains after the beamforming.

\section{Probabilistic formulation}
\newcommand{\boldmu}{\mbox{\boldmath $\mu$}}

Let $\theta_w=\{\mathbf{w}_0,\mathbf{w}_b,\ldots,\mathbf{w}_{L_w}\}$, $\theta_B=\{B_0, B_b,\ldots,B_{L_w}\}$, $\theta_{\sigma}=\{\sigma_t^2\mid t=1,2,\ldots,T\}$, and $\theta=\{\theta_w,\theta_B,\theta_{\sigma},\mathbf{v}\}$ be model parameter sets, where $\sigma_t^2$ is the time-varying power of the desired signal, and $T$ is the number of available time frames. Then, assuming based on Eq.~(\ref{eq:bf}) that the probabilistic uncertainty of $\mathbf{x}_t$ is derived only from $\mathbf{y}_t$ when $\mathbf{x}_{t-\tau}$ for $\tau=b,b+1,\ldots,L_w$ are given, the likelihood function can be defined and rewritten as
\begin{align}
  {\cal L}(\theta)&=\log p(\{\mathbf{x}_t\};\theta),\\
  &=\sum_t \log p(\mathbf{x}_t\mid\mathbf{x}_{t-b},\mathbf{x}_{t-b-1},\ldots,\mathbf{x}_{t-L_w};\theta)\label{eq:condprob}\\
  &=\sum_t \log p\left(W_0^{\HT}\mathbf{x}_t\mid\sum_{\tau=b}^{L_w}W_\tau^{\HT}\mathbf{x}_{t-\tau};\theta\right)\nonumber\\
  &\hspace{4mm}+2T\log|\det(W_0)|,\\
  &=\sum_t \log p(\mathbf{y}_t;\theta)+2T\log|\det(W_0)|.\label{eq:like2}
\end{align}

Now, we assume that the optimal beamformer $\theta_w$ can reduce the level of the reverberation $\tilde{r}_t$ and that of the noise $\tilde{n}_t$ in Eq.~(\ref{eq:output2}) to be negligibly small, i.e., $\tilde{r}_t+\tilde{n}_t\simeq 0$. Then, the first row and the remaining rows in Eqs.~(\ref{eq:output2}) can be considered statistically independent of each other, and thus $p(\mathbf{y}_t;\theta)$ can be decomposed into $p(y_t^{(1)};\theta_w)$ and $p(\mathbf{y}_t^{(2:M)};\theta_B)$.
Then, the likelihood function can be rewritten as
\begin{align}
  {\cal L}(\theta)&=\sum_t\log p(y_t^{(1)};\theta_w,\theta_{\sigma})+\sum_t\log p(\mathbf{y}_t^{(2:M)};\theta_B)\nonumber\\
  &\hspace{4mm}+2T\log\frac{|v^{(1)}|}{||\mathbf{v}||_2}+T\log\det(B_0^{\HT}B_0),\label{eq:like3}
\end{align}
where the last two terms in Eq.~(\ref{eq:like3}) are derived from the last term in Eq.~(\ref{eq:like2}) based on Eqs.~(\ref{eq:dr}) and (\ref{eq:orth}) (See Appendix), and $||\cdot||_2$ denotes the Euclidean norm of a vector. 

Note that because it is difficult to optimize $\mathbf{v}$ based on the ML estimation, we estimate it separately from the ML estimation as described in Section~\ref{sec:v}.  Then, for the estimation of $\theta_w$, we need only to estimate $\Theta=\{\theta_w,\theta_{\sigma}\}$, based solely on the first term in Eq.~(\ref{eq:like3}).

Finally, we introduce a model of the probability density function (pdf) of $y_t^{(1)}$ as
\begin{align}
  p(y_t^{(1)};\theta_w,\theta_{\sigma})&={\cal N}_C(y_t^{(1)}=\bar{\mathbf{w}}^{\HT}\bar{\mathbf{x}}_t;0,\sigma_t^2), \label{eq:pdf}
\end{align}
where ${\cal N}_{C}({x};\mu,\sigma^2)$ is a pdf of a complex Gaussian distribution with a mean $\mu$ and a covariance $\sigma^2$, defined as
\begin{equation}
  {\cal N}_{C}({x};\mu,\sigma^2)=\frac{1}{\pi\sigma^2}\exp\left(-\frac{|{x}-\mu|^2}{\sigma^{2}}\right).
\end{equation}
Then, 
the likelihood function to be maximized becomes
\begin{align}
  {\cal L}(\Theta)&=-\sum_t\frac{|y_t^{(1)}|^2}{\sigma_t^2}\label{eq:like4.0}-\sum_t\log \sigma_t^2,\\
  &=-\sum_t\frac{|\bar{\mathbf{w}}^{\HT}\bar{\mathbf{x}}_t|^2}{\sigma_t^2}-\sum_t\log \sigma_t^2,\label{eq:like4}
\end{align}
where we set $\bar{\mathbf{w}}=[\mathbf{w}_0^\top,\mathbf{w}_b^\top,\mathbf{w}_{b+1}^\top,\ldots,\mathbf{w}_{L_w}^\top]^\top$ and $\bar{\mathbf{x}}_t=[\mathbf{x}_{t}^\top,\mathbf{x}_{t-b}^\top,\mathbf{x}_{t-b-1}^\top,\ldots,\mathbf{x}_{t-L_w+1}^\top]^\top$.
Note that $\bar{\mathbf{w}}$ and $\bar{\mathbf{x}}_t$ have a time gap between their first and the second elements, corresponding to the prediction delay $b$.

\section{Solution to ML estimation}
Because it is difficult to obtain a closed form solution that maximizes Eq.~(\ref{eq:like4}), we adopt an iterative estimation scheme, by which Eq.~(\ref{eq:like4}) is maximized to a stationary point by alternately updating $\theta_w$ and $\theta_{\sigma}$ from certain initial values.  
  
In the step employed to update $\theta_w$, while fixing $\hat{\sigma}_t^2$ for each $t$, $\theta_w$ is updated as one that maximizes the likelihood function shown below with the distortionless constant in Eq.~(\ref{eq:dr}). 
\begin{align}
  {\cal L}(\theta_w)&=-\sum_t\frac{|\bar{\mathbf{w}}^{\HT}\bar{\mathbf{x}}_t|^2}{\hat{\sigma}_t^{2}}
~\mathrm{s.t.}~\mathbf{w}_0^{\HT}\mathbf{v}=v^{(1)}, \label{eq:wmpdrest}
\end{align}
The solution that maximizes Eq.~(\ref{eq:wmpdrest}) can be derived based on the Lagrange multiplier method as
\begin{equation}
\hat{\bar{\mathbf{w}}}=\frac{{R}^{-1}\bar{\mathbf{v}}}{\bar{\mathbf{v}}^{\HT}{R}^{-1}\bar{\mathbf{v}}},\label{eq:west}
\end{equation}
where $\bar{\mathbf{v}}=[\mathbf{v}^\top/v^{(1)},0,0,\ldots,0]^\top$ is a column vector containing $\mathbf{v}/v^{(1)}$ followed by $M(L_w-b+1)$ zeros, and $R$ is a power-normalized temporal-spatial covariance matrix with a prediction delay calculated as
\begin{equation}
{R}=\sum_t\frac{\bar{\mathbf{x}}_t\bar{\mathbf{x}}_t^{\HT}}{\hat{\sigma}_t^{2}}.\label{eq:Rest}
\end{equation}
Then, the estimate of the desired signal is obtained as
\begin{equation}
  \hat{d}_t^{(1)}=\hat{\bar{\mathbf{w}}}^{\HT}\bar{\mathbf{x}}_t.\label{eq:dest}
\end{equation}
%


In the other step used to update $\theta_{\sigma}$, $\sigma_t^2$ can be updated as the power of the estimated desired signal, i.e., $\hat{\sigma}_t^2=|\hat{d}_t^{(1)}|^2$. 

It may be worth noting that the ML formulation presented here is reduced to that for MPDR (and its realization based on a generalized sidelobe canceller) if we set $\mathbf{r}_t=0$ for all $t$ and $W_{\tau}=0$ for $\tau\ge b$ in the formulation and assume that $\sigma_t^2$ is time invariant.  

\subsection{Analysis of the solution}
Because $d_t^{(1)}$, $\tilde{r}_t$, and $\tilde{n}_t$ in Eq.~(\ref{eq:output2}) are mutually independent, Eq.~(\ref{eq:wmpdrest}) can be decomposed, under the distortionless constraint, as
\begin{align}
  {\cal L}(\theta_w)
  &=-\sum_t\frac{|d_t^{(1)}|^2}{\hat{\sigma}_t^2}-\sum_t\frac{|\tilde{r}_t|^2}{\hat{\sigma}_t^2}-\sum_t\frac{|\tilde{n}_t|^2}{\hat{\sigma}_t^2}.\label{eq:decmp2}
\end{align}
The first term in Eq.~(\ref{eq:decmp2}) does not depend on $\theta_w$. Thus if we obtain a beamformer $\theta_w$ that maximizes Eq.~(\ref{eq:wmpdrest}) for fixed $\hat{\sigma}_t^2$, it surely maximizes the sum of the second and third terms in Eq.~(\ref{eq:decmp2}).  As a consequence, the beamformer can perform denoising and dereverberation simultaneously.  

\subsection{Overall processing flow with estimation of $\mathbf{v}$ using WPE}
\label{sec:v}
\begin{figure}[!t]
\centering
\includegraphics[width=3.2in]{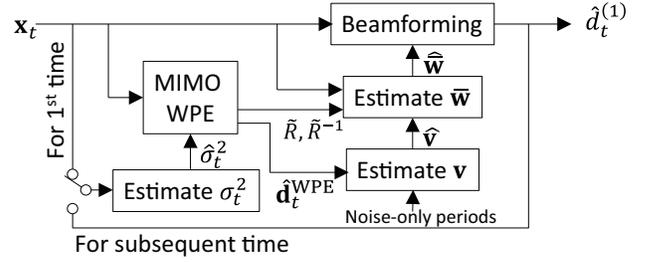}
\caption{Iterative beamformer estimation by WPD performed with WPE.}
\label{fig_sim}
\end{figure}

For the accurate estimation of $\mathbf{v}$ from the captured signal, it is crucial to exclude any influence of noise and reverberation.  For this purpose, we first reduce the effect of reverberation by performing MIMO dereverberation using WPE \cite{gwpe} within the WPD framework (see Fig.~\ref{fig_sim}).
Because WPE and WPD share most of the calculation that requires the majority of their computing cost, namely the calculation of $R$ in Eq.~(\ref{eq:Rest}) and its inverse, WPE can be performed very efficiently within the  WPD framework to obtain the dereverberated multichannel signal, $\hat{\mathbf{d}}_t^{\mbox{\footnotesize WPE}}$.  In concrete terms, let $\tilde{R}$ be an $M(L_w-b+1)\times M(L_w-b+1)$ dimensional submatrix of $R$, obtained by excluding the first $M$ rows and $M$ columns from $R$.  For WPE, we need to calculate $\tilde{R}$ and its inverse, which accounts for the majority of the computing cost of WPE.  Then, once we have $\tilde{R}$ and its inverse, we can calculate $R$ and its inverse with little additional computing cost \cite{2x2inverse}. 
Finally, we estimate $\mathbf{v}$ from $\mathbf{d}_t^{\mbox{\footnotesize WPE}}$ by reducing the effect of the noise based on generalized eigenvalue decomposition with noise covariance whitening \cite{ito17icassp,Gannot15icassp}, assuming that the noise-only periods are given.  


Figure~\ref{fig_sim} shows the overall processing flow of WPD when utilizing WPE to estimate $\mathbf{v}$.  Adopting the power of the captured signal as the initial value of $\sigma_t^2$, we iterate WPD jointly with WPE, and update $\mathbf{v}$ and $\sigma_t^2$ using the outputs of WPE and WPD, respectively.  In each iteration, the beamformer $\hat{\bar{\mathbf{w}}}$ and the desired signal $\hat{d}_t^{(1)}$ are updated based on Eqs.~(\ref{eq:west}), (\ref{eq:Rest}), and (\ref{eq:dest}) using the estimated $\hat{\mathbf{v}}$ and $\hat{\sigma}_t^2$.

\section{Experiments}
\subsection{Dataset and evaluation metrics}
We evaluated the performance of the proposed method using the REVERB Challenge dataset \cite{REVERB}.
The evaluation set (Eval set) of the dataset is composed of simulated data (SimData) and real recordings (RealData). 
Each utterance in the dataset contains reverberant speech uttered by a speaker and stationary additive noise. The distance between the speaker and the microphone array is varied from 0.5~m to 2.5~m. 
For SimData, the reverberation time is varied from about 0.25 s to 0.7 s, and the signal-to-noise ratio (SNR) is set at about 20 dB.

Evaluation metrics prepared for the challenge were used in the experiments. As objective measures for evaluating speech enhancement performance \cite{metrics}, we used the cepstrum distance (CD), and the frequency-weighted segmental SNR (FWSSNR). To evaluate the ASR performance, we used a baseline ASR system recently developed using Kaldi \cite{kaldi}. This is a fairly competitive system composed of a TDNN acoustic model trained using lattice-free MMI and online  i-vector extraction, and a tri-gram language model.

\subsection{Methods to be compared and analysis conditions}
\begin{table}[!t]
\renewcommand{\arraystretch}{1.2}
\caption{CD (dB), FWSSNR (dB), and WER (\%) of enhanced speech obtained after 1st iteration using REVERB Challenge eval set. No Enh means no speech enhancement. Boldface indicates the best score for each metric. }\label{tbl:aq}
\label{table_example}
\centering
\begin{tabular}{|c|c|c|c|c|}
\hline
  & \multicolumn{3}{c|}{SimData} & RealData\\
\cline{2-5}
       & CD & FWSSNR & WER & WER\\\hline
No Enh   & 3.97 & 3.62 & 4.35 &  18.61 \\
MPDR     & 3.43 & 5.97 & 5.56 &  14.68\\
WPE      & 3.74 & 4.79 & 4.37 &  13.44\\
WPE+MPDR & 3.02 & 7.30 & 4.42 &  10.46\\
WPD w/o WPE & 3.23 & 6.25 & 4.82 & 12.06 \\
WPD w/ WPE & {\bf 2.65} & {\bf 7.98} & {\bf 3.83} & {\bf 9.90} \\
\hline
\end{tabular}
\end{table}
WPD (Proposed) was compared with WPE, MPDR, and the integration of WPE followed by MPDR in a cascade configuration (WPE+MPDR).  To confirm the importance of utilizing WPE within the WPD framework, we examined the performance of WPD with and without WPE. The two configurations are respectively referred to as WPD w/ and w/o WPE.  Without WPE, WPD estimates $\mathbf{v}$ from the captured signal, and does not update it during the iterative estimation.

For all the methods, a Hann window was used for a short-time analysis with the frame length and the shift set at 32~ms and 8~ms, respectively. The sampling frequency was 16 kHz and $M=8$ microphones were used for all the experiments. For WPE, WPE+MPDR, and WPD, the prediction delay was set at $b=4$, and the length of the prediction filter was set at $L_w=12, 10$, and $6$, respectively, for frequency ranges of $0$ to $0.8$ kHz, $0.8$ to $1.5$ kHz, $1.5$ to $8$ kHz.  For the estimation of $\mathbf{v}$, we assumed that each utterance had noise-only periods of 225 ms and 75 ms, respectively, at its beginning and ending parts.

\subsection{Evaluation results}
Table~\ref{tbl:aq} summarizes the CDs, FWSSNRs, and WERs of the captured signals and the enhanced signals obtained after the first estimation iteration.  In the table, all the methods improved the captured signal with all the measures except for the WERs on SimData. While WPD w/ WPE performed the best of all, WPD w/o WPE did not perform very well.  This indicates that the reliable estimation of $\mathbf{v}$ is very important for successful beamforming by WPD, and it can be achieved by utilizing WPE for the estimation of $\mathbf{v}$ within the WPD framework.

\begin{figure}[!t]
  \centering
  \includegraphics[width=3.3in]{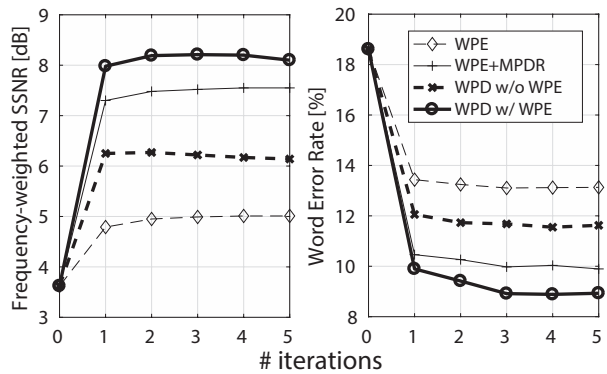}
  \caption{Performance curve of FWSSNRs (dB) and WERs (\%) with increases in \# of estimation iterations.  FWSSNRs are evaluated using SimData of Eval set and WERs are evaluated using RealData of Eval set.}\label{fig:pcurve}
\end{figure}

Figure~\ref{fig:pcurve} shows the performance curve of the methods in terms of FWSSNRs and WERs when we performed the iterative estimation. We confirmed that WPD with WPE again greatly outperformed all the other methods for all the iteration times, and the iterative estimation was effective at least in the first few steps for WPD.

\section{Concluding remarks}
We presented a probabilistic formulation of WPD that achieves dereverberation and denoising both simultaneously and optimally, and derived an optimization algorithm based on the ML estimation.  Furthermore, we proposed a method for effectively estimating the steering vector of the desired signal by incorporating WPE into the WPD framework.  The experiments showed that WPD with WPE greatly outperformed conventional approaches, including one that utilizes WPE followed by MPDR in a cascade configuration, and that the incorporation of WPE into the WPD framework is very important if we are to achieve a reliable estimation of the steering vector for successful beamforming.

\appendix[Decomposition of $|\det(W_0)|$]
Let us decompose the first column of $W_0$ in Eq.~(\ref{eq:Wt}) as $\mathbf{w}_0=\mathbf{e}+\mathbf{b}$, where $\mathbf{e}$ is a projection of $\mathbf{w}_0$ to $\mathbf{v}$, which is determined based on Eq.~(\ref{eq:dr}) as
\begin{align}
  \mathbf{e}&=\frac {(v^{(1)})^{\HT}}{||\mathbf{v}||_2^2}\mathbf{v},\label{eq:ev}
\end{align}
and $\mathbf{b}$ is a component that is orthogonal to $\mathbf{v}$. Note that $\mathbf{b}$ is linearly dependent on the subspace spanned by $B_0$.

Then, $|\det(W_0)|$ can be expanded as
\begin{align}
  |\det(W_0)|&=|\det([\mathbf{e}+\mathbf{b},B_0])|,\\&=|\det([\mathbf{e},B_0])+\det([\mathbf{b},B_0])|,\\&=|\det([\mathbf{e},B_0])|,\\
  &=\det\left(\left[\begin{array}{c}\mathbf{e}^{\HT}\\B_0^{\HT}
\end{array}\right][\mathbf{e},B_0]\right)^{1/2},\\
  &=\det\left(\left[\begin{array}{cc}
      \mathbf{e}^{\HT}\mathbf{e} & 0\\
      0 & B_0^{\HT}B_0
    \end{array}\right]\right)^{1/2},\\
  &=\det(\mathbf{e}^{\HT}\mathbf{e})^{1/2}\det(B_0^{\HT}B_0)^{1/2},\\
  &=\frac{|v^{(1)}|}{||\mathbf{v}||_2}\det(B_0^{\HT}B_0)^{1/2}.\label{eq:logdetw}
\end{align}
As a consequence, $|\det(W_0)|$ is decomposed into one based only on $\mathbf{v}$ and another based only on $B_0$.  Note that $\det(B_0^{\HT}B_0)^{1/2}$ is equal to the absolute value of the product of all the singular values of $B_0$.

  \section*{Acknowledgment}

We would like to thank Mr. Christoph Boeddeker and Prof. Reinhold Haeb-Umbach for valuable comments on the equivalence between cascaded and unified optimization approaches of the WPD.

\bibliographystyle{IEEEtran}
\bibliography{bibs}

\end{document}